\documentclass[twocolumn,pra,showpacs,preprintnumbers,amsmath,amssymb]{revtex4}

\usepackage{amssymb}

\usepackage{graphics}

\usepackage{epsfig}

\usepackage{graphicx}    

\usepackage{dcolumn}     

\usepackage{amsthm}

\usepackage{bm}          

\usepackage{amsbsy}

\begin{document}

\title{Electronic Structure in Gapped Graphene with Coulomb Potential}

\author{W. Zhu$^{1}$, M. L. Liang$^2$, Q. W. Shi$^{1 \dagger}$, Z. F. Wang$^1$, J. Chen$^3$,$^4$, J. G. Hou$^{1}$}

\address{$^1$Hefei National Laboratory for Physical Sciences at
Microscale, University of Science and Technology of China, Hefei
 230026, China}
\address{$^2$Department of Physics,University of Science and
Technology of China, Hefei 230026, China}
\address{$^3$Electrical and Computer Engineering, University of Alberta, Alberta, Canada T6G 2V4}
\address{$^4$National Research Council/National Institute of
Nanotechnology, Alberta, CANADA T6G 2M9}
\email[Electronic
address:]{phsqw@ustc.edu.cn}

\date{\today}

\begin{abstract}
In this paper, we numerically study the bound electron states
induced by long range Coulomb impurity in gapped graphene and the
quasi-bound states in supercritical region based on the lattice
model. We present a detailed comparison between our numerical
simulations and the prediction of the continuum model which is
described by the Dirac equation in ($2+1$)-dimensional Quantum
Electrodynamics (QED). We also use the Fano's formalism to
investigate the quasi-bound state development and design an
accessible experiments to test the decay of the supercritical
vacuum in the gapped graphene.
\end{abstract}

\pacs{81.05.Uw, 71.55.-i, 71.23.-k}

\maketitle

\textit{\bf Introduction} Graphene, a two-dimension (2D) hexagonal
lattice of carbon atoms, exhibits special electronic dispersion
relation, in which electrons behave like massless relativistic
Dirac fermions \cite{a1,a2,a3}. This special property leads to
many unconventional phenomena, such as the Klein paradox\cite{a10}
and the Vaselago lensing effect\cite{a11}. Due to its large "fine
structural constant", graphene not only provides an exciting
platform to validate some predictions of  Quantum Electrodynamics
(QED) in the strong field,  but also provides an interesting
"strong coupling" 2+1 QED model\cite{a12}.

Recently, it was found that a substrate induced potential can
break the chiral symmetries of massless Dirac equation and
generate gaps in graphene electron spectrum\cite{b1}. The motion
of electrons, therefore, can be described by massive 2D Dirac
equations. It was predicted that a Coulomb charged impurity in
gapped graphene behaves the same as the heavy-atom in the QED
theory. The charged impurities have a significant effect on the
electronic structures of the gapped graphene
\cite{a15,a16,a17,a18,a19}. Depending on the charge $Z$, the bound
states can be induced inside the gap, and the quasi-bound states
can also be generated when the charge $Z$ is sufficiently larger
than the supercritical number $Z_c$. If the energy of bound state
is above the middle of the gap, the wave function of bound
electron state in the pure Coulomb potential can be well
determined by using the continuum model. If the energy of bound
state is below the middle of the gap, the wave function of bound
state, however, cannot be well described because of the singular
nature of the pure Coulomb potential near $r\sim 0$. To solve this
problem, a rough but simple approximation is adopted to remove
this singularity condition by taking into account the finite
spatial extension of nuclear charge with radius $R$ based on 3+1
QED model \cite{b4}. When a continuum model is used to describe
the behavior of electrons in graphene near the Dirac point, in
fact, how to choose the boundary condition $r\sim0$ is still an
open question. For example, Ref.\cite{a40} chose a ``Zigzag edge"
boundary condition to discuss the vacuum polarization in gapless
graphene, but Ref.\cite{a41} used an ``infinite mass" boundary
condition. This situation, however, dose not exist in the
tight-binding model due to its lattice scale cutoff.

In this paper, based on the lattice model, we use the tight-binding
approach to numerically calculate the electronic structure in gapped
graphene.  From the results of local density of states (LDOS), we
discover that the energy position of bound states can be determined
within the non-supercritical region, and supercritical charge $Z_c$
can also be determined using the continuum model and the lattice
model. We also discuss the relationship between the gap and critical
charge $Z_c$. Finally, we use the description of resonances to
explain the quasi-bound state development in a supercritical regime.

\textit{\bf The Lattice Model and Continuum Model} In what follows,
we consider a single attractive Coulomb impurity placed at the
center of a honeycomb lattice in a graphene sheet. The corresponding
Hamiltonian in the tight-binding form is:

\begin{eqnarray}
H=-t(\sum_{i,j}{a^\dagger_i b_j +h.c.})+M\sum_{i}{(a^\dagger_i
a_i-b^\dagger_i b_i)}\nonumber
\\-\frac{Ze^2}{\varepsilon}\sum_{i}{(\frac{a^\dagger_i a_i}{r^{A}_i}+\frac{b^\dagger_i b_i}{r^{B}_i})},
\end{eqnarray}

where $t=2.7eV$ is hopping energy between the nearest neighbor
atoms. The operator $a^\dagger$($a$) and $b^\dagger$($b$) denote
creation (annihilation) an electron on the sublattice A and
sublattice B, respectively. The first term of Hamiltonian
describes the hopping between the nearest neighbor atoms. The
second term is the on-site energy of sublattice A and sublattice
B. $M$ corresponds to the mass (or the gap) of Dirac fermions. It
is well known that the non-zero mass term can open a $2M$ wide gap
within the band spectrum. Many conditions can lead to a band-gap
within graphene. For instance, Ref\cite{b1} reports that the SiC
substrate can open a gap of $\sim 0.26eV$ in a single-layer
graphene, which is consistent with the calculation of the first
principle \cite{b2}. In our calculations, we select $M$ from
$0.05t$ (or $0.13$eV) to $0.10t$ (or $0.27$eV). $Z$ reflects the
impurity charge strength and $\varepsilon$ is the effective
dielectric constant.  When energy level is close to the Dirac
point, the Hamiltonian in gapped graphene can be approximately
described by a continuum model:

\begin{eqnarray}
H=-iv_F(\sigma_1\partial_x+\sigma_2\partial_y-\frac{Z\alpha}{r})+M\sigma_3
\end{eqnarray}

where $\sigma_{1,2,3}$ are the Pauli matrices. We can view
$\alpha=\frac{e^2}{\varepsilon \hbar v_F}$ as ``fine structural
constant" in graphene. $v_F=\frac{3}{2}ta$ is the Fermi velocity
of graphene and $a=1.42{\AA}$. $M$ is the mass of the Dirac
fermion. Since $v_F$ is sufficiently small compared with the
velocity of light in the QED theory,  $\alpha\sim 1$ then becomes
much larger than that in the QED theory. The large value of "fine
structural constant" of graphene means that the perturbation
expansion, which works well in QED, can not directly apply to
discuss the many body problems in graphene .

\begin{figure}
\includegraphics[width=0.5\textwidth]{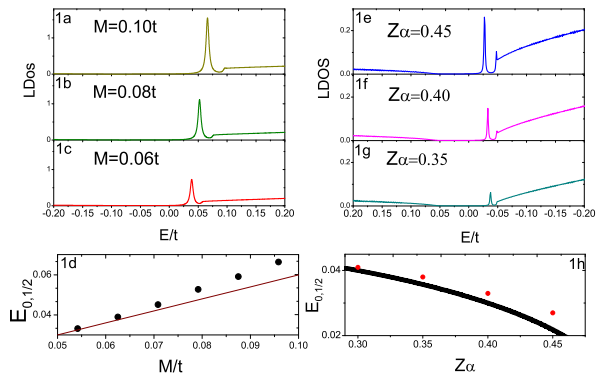}
\caption{(Color online)Subfigures 1a,1b and 1c describe the bound
states of the nearest neighbor with different gaps. 1d shows the
relationship between the energy of bound states and the gap when
$Z\alpha=0.4$. The line is the theoretical prediction for
$Z\alpha=0.4$ and $j=1/2$. Subfigures 1e,1f and 1g describe the
bound states of the nearest neighbor with different Coulomb charge
value when $M=0.05t$. Subfigure 1h shows the relationship between
the energy of bound states and the Coulomb charge values when
$M=0.05t$. The black line is the theoretical prediction.}
\end{figure}

\textit{\bf Bound states above the middle gap} In the continuum
model, the eigenfunction of a pure Coulomb potential can be
described by the confluent hypergeometric function \cite{a40,a50}.
The regularity at $r\rightarrow0$ and $r\rightarrow\infty$ requires
the confluent hypergeometric functions reduce to polynomials at the
same time. Energy of the bound states within the gap $|E|<M$, as a
result, satisfies the following condition:

\begin{eqnarray}
E_{n,j}=\frac{M\emph{sgn}(Z\alpha)}{\sqrt{1+\frac{(Z\alpha)^2}{(n+\gamma)^2}}}
\end{eqnarray}

with the root $\gamma=\sqrt{j^2-(Z\alpha)^2}$. Here $n$ and $j$
takes an integer value and half of an integer, respectively. That
is, $n=0,1,2..$ and $j=1/2, 3/2..$. $j$ is the isospin-orbital
momentum number \cite{a50}. Non-supercritical region is defined
because $\gamma$ is always real for all angular momentum channels.
If the control charge exceeds the value $Z\alpha>j$, the solutions
of the continuum model break down since the root $\gamma$ becomes
imaginary. Hence, all bound states below the middle of gap do not
exist. This artifact can be remedied by removing the singular
behavior of the pure Coulomb potential. When the energy of the bound
state above the middle of gap, one can see from Eq. $(3)$ that the
energy of the bound state is linearly proportion with the gap width,
and the lowest bound state reaches the middle of the gap $E=0eV$ at
a critical value that satisfies $Z\alpha=j=1/2$.

\begin{figure}
\includegraphics[width=0.45\textwidth]{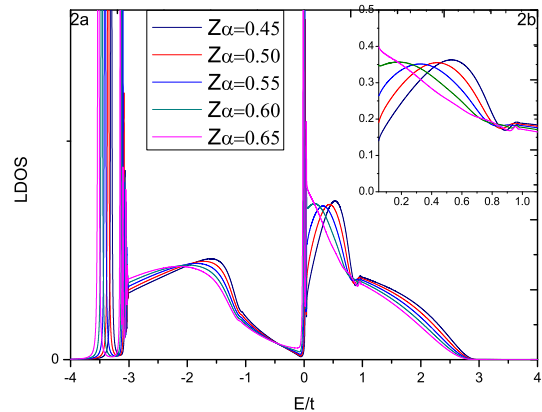}
\caption{(Color online)2a. LDOS of the nearest neighbor atoms with
different $Z\alpha$) when the gap $M$ is set to be $0.05t$. 2b. LDOS
at the energy scale $M<E<t$ differs obviously when the bound state
reaches $0eV$.}
\end{figure}

As shown in Fig.1, the LDOS spectrum for the bound state above the
middle gap is calculated numerically by the tight-binding approach.
By fixing the electromagnetic coupling $Z\alpha=0.4$, we can plot
the LDOS spectrum of the bound states with different gap widthes
($M$) as shown in Fig. 1a to Fig. 1c. Except that the lowest bound
states ($n=0$ and $j=1/2$) in the gap can be clearly resolved based
on the LDOS spectrum, the other bound states are very close to the
edge of the positive energy continuum and thus one cannot clearly
distinguish them from Fig.1a to Fig.1c. The energy of the lowest
bound state is plotted in Fig.1d as a function of the gap width. The
dotted line is depicted based on our numerical simulations in the
lattice model. The theoretical results is also plotted by using the
equation $(3)$ of the continuum model. When the gap width is not too
large, the energy of the lowest bound state calculated by the
numerical simulation show a slightly deviation away with that
predicted by the continuum model. As the gap $M$ increases, the
deviation becomes larger.

If the gap width is set to be $M=0.05t$, Fig1.e to Fig.1g illustrate
the LDOS spectrum of the bound states with different chosen
electromagnetic coupling factor, $Z\alpha$.  The peak of LDOS
corresponding to the lowest bound state gets close to the middle gap
as $Z\alpha$ increases. Fig. 1h plots the energy of the lowest bound
state as a function of $Z\alpha$. The dotted line gives the result
of numerical simulations and the fitting curve is plotted using
equation (3) of the continuum model. The numerical simulations fit
Eq. ($3$) well when $Z\alpha$ is small. Fig. 1h also imply that the
electromagnetic coupling $Z\alpha$ needs to be larger than
$Z\alpha=1/2$ in order to make the energy of the lowest bound state
to reach the middle of the gap.

The effect of Coulomb charge on energy continuum is shown in
Fig.2. Except for the existent of bound states in the gap, the
main feathers of the LDOS spectrum on gapped graphene is almost
the same as that in the gapless graphene \cite{a41}.  The strong
re-normalization of the van Hove singularities are also
observed.\cite{a41}. Note that the LDOS spectrum near the edge of
positive energy continuum appears some interesting physical
signature, which is discussed in the following.

\textit{\bf Bound states below the middle gap} The Dirac equation
with the pure Coulomb potential can only provide the solution of
bound state above the middle of gap. When $Z\alpha$ exceeds
$j=1/2$, $\gamma$ becomes imaginary in this continuum model, which
does not have a regular solution. For the pure Coulomb potential,
the continuum model points out that this is critical point when
$Z\alpha$ equals $1/2$. In the lattice model, the bound state,
however, can be gradually enter the energy region below the middle
of gap as $Z\alpha$ increases. Thus, this point seems not
critical. In this paper, for a convenient comparison between the
results obtained by the lattice model and by the continuum model,
we still define a critical value $Z_0\alpha$ which represents the
energy of the lowest bound state just touches the middle of gap
$E=0eV$.  More interesting, our numerical simulations reflect that
some physical signatures are associated to this point.

Fig.3b $\sim$ Fig.3d show the situation that the bound states
enter below the middle of gap when we fix the gap width $M=0.05t$.
The bound states exist below the middle gap as the $Z\alpha$
increases. The relationship between the critical charge
$Z_0\alpha$ and the gap width $M$ is numerically calculated  and
the results are shown in Fig.3a. The trigonal dots stands for the
numerical results of critical charge $Z_0\alpha$. Therefore, the
value $Z_0\alpha$ depends on the gap width $M$ in our lattice
model. For example, the critical value $Z_0\alpha$ is about $0.6$
when the gap width is $M=0.05t$. It is more interesting to note
that there still exist some physical signatures related to this
critical value.  As shown in Fig.2b, the LDOS spectrums of the
positive energy continuum near the gap appear significantly
different for $Z\alpha<Z_0\alpha$ versus $Z\alpha>Z_0\alpha$.
When $Z\alpha<Z_0\alpha$, the LDOS of positive energy continuum
decreases as $E/t$ gets close to the gap region. However, for
$Z\alpha=Z_0\alpha$, the LDOS of positive energy continuum near
the gap is almost independent of $E/t$. When $Z\alpha>Z_0\alpha$,
the LDOS spectrum near the gap appears an increasing trend as
$E/t$ decreases. The physics behind this phenomenon needs to be
explained in our future research.

\begin{figure}
\includegraphics[width=0.45\textwidth]{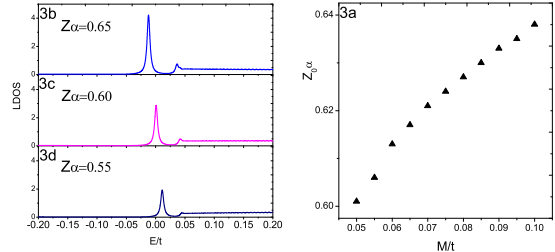}
\caption{(Color online)3a.Relationship between critical charge
$Z_0\alpha$ and Dirac mass. Subfigures 3b, 3c and 3d describe the
bound states for $Z\alpha=0.55,0.60$ and $0.65$ in the gap,
respectively.}
\end{figure}

\textit{\bf Supercritical Region and the Decay of Vacuum} As
$Z\alpha$ continues to increase, the energy of the lowest bound
state becomes close to the edge of band gap ($E=-M$). When
$Z\alpha$ is sufficiently larger than the critical value
$Z_c\alpha$, the energy of bound state will drop into the negative
continuum spectrum. In this case,  the bound state changes its
characteristics and becomes a quasi-bound state. As predicted by
the QED theory\cite{b4}, the neutral vacuum changes into the
charged vacuum. The processes with this change is called the decay
of vacuum. We first calculate the supercritical charge $Z_c\alpha$
in gapped graphene by using the continuum model and the lattice
model, respectively. To determine the diving point by the
continuum model, one needs to remove the singular behavior of the
pure Coulomb potential. A simple choice is that the potential
takes the form $V=-\frac{Z\alpha}{R},r<R$, and
$V=-\frac{Z\alpha}{r}$, otherwise. Hence, the supercritical charge
$Z_c\alpha$ for the lowest bound state can be determined through
the boundary condition at $r=R$\cite{b4,b5}:

\begin{eqnarray}
\frac{J_1(Z_c\alpha)}{J_0(Z_c\alpha)}=\frac{1}{2Z_c\alpha}(1-\rho_c\frac{K'_{i\nu}(\rho_c)}{K_{i\nu}(\rho_c)})
\end{eqnarray}

where $\rho_c=\sqrt{8MRZ_c\alpha}$ and
$\nu=2\sqrt{(Z_c\alpha)^2-j^2}$. This is a transcendental equation
for $Z\alpha$ and the numerical result gives $Z_c\alpha=0.78$ at
$MR=0.02$. If the gap width is set to be $M=0.05t$,  the
corresponding $R$ equals about $0.6a$.

Based on our lattice model, Fig.4a shows the numerical results when
the lowest bound state enters into the negative energy continuum.
The results indicate that the supercritical charge is about
$Z_c\alpha=0.756$ when $M=0.05t$. Inset in fig.4a reflects the
relationship between $Z_c\alpha$ and the gap width $M$. The square
dots represents the numerical results based on the lattice model.
The fitting line implies that $Z_c\alpha$ depends linearly on the
gap width $M$, which satisfies the result of the above
transcendental equation $(4)$, when $MR<<1$.

\begin{figure}
\includegraphics[width=0.45\textwidth]{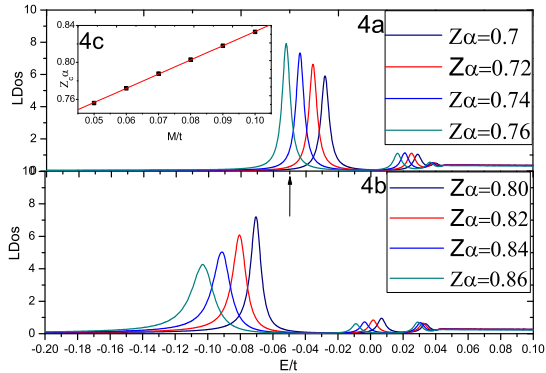}
\caption{(Color online)4a. The evolving bound states near $E=-M$
when $Z\alpha<Z_c\alpha$. We can observe that LDOS on the bound
states becomes larger. 4b. The evolving resonances near $E=-M$ when
$Z\alpha>Z_c\alpha$. The arrow points towards the edge of gap
$E=-M$. The inset subfigure 4c shows the relationship between
$Z_c\alpha$ and gap width $M$.}
\end{figure}

If $Z\alpha>Z_c\alpha$, the lowest bound state enters the negative
energy continuum and becomes a resonant state. The LDOS spectrum
clearly shows the resonant state in negative energy continuum (refer
to Fig.4b). The LDOS spectrum also shows that the second lowest
bound state is distinguishable. From the numerical results, we can
see that the width of quasi-bound state becomes large as $Z\alpha$
increases. This behavior can be explained by the method developed by
U. Fano for the auto-ionization of excited states in Atomic Physics.
We assume that new continuum wave function in the supercritical
region can be expanded as $|\chi_E>=a(E)|\phi_0>+\int
dE'b_{E'}(E)|\psi_{E'}>$\cite{a60}, where $|\phi_0>$ and $|\psi_E>$
represent the discrete eigenstate and the continuous spectrum when
$Z\alpha=Z_c\alpha$. If $Z\alpha$ exceeds $Z_c\alpha$,
$V=(Z\alpha-Z_c\alpha)/r=\delta_{Z\alpha}/r$ can be considered as a
perturbation potential. On the boundary of the supercritical region,
we can simply revise the results in Ref.\cite{a60}:

\begin{eqnarray}
|a(E)|^2=\frac{\eta/2\pi}{(E-E_0-\Delta E)^2+\eta^2/4}
\end{eqnarray}

where $\Delta E=\delta_{Z\alpha}<\phi_0|1/r|\phi_0>$ and
$\eta=2\pi|V_E|^2=2\pi(\delta_{Z\alpha})^2<\varphi_E|1/r|\phi_0>$.
The LDOS is defined as
$N(\epsilon,r)=\sum_E|<r|\chi_E>|^2\delta(\epsilon-E)$\cite{a40}. We
obtain that $N(\epsilon,r)$ relates to $|\phi_0>$ as
$|a(E)|^2|<r|\phi_0>|^2$. $E_r=E_0+\Delta E$  depends linearly on
additional Coulomb charge while  width of the quasi-bound $\eta$
quadratically increases with the additional Coulomb charge. These
observations is consistent with our numerical results qualitively as
shown in fig.4b.

Particularly important, some gedanken experiments were proposed to
test the interesting physical process related to the supercritical
vacuum.\cite{b4} For example, the feature of the spectrum of the
emitted positron depending on the "critical duration" which
reflects the decay of the supercritical vacuum. Unfortunately,
there still exist difficulty to perform this experiment in QED.
Since ``the Compton wavelength $\lambda_g=\hbar v_F/M$" in gapped
graphene is sufficiently large comparing with that in QED, one can
use the gating technique to design experiments to measure the
decay of the supercritical vacuum in a gapped graphene.


\textit{\bf Summary} In this paper, considering the contribution
of the long-range Coulomb impurity in gapped graphene, we
numerically calculate the LDOS spectrum using the lattice model.
The relationship between the critical value $Z\alpha$ and gap $M$
is investigated, which corresponds to the lowest bound state at
the middle gap or at the diving point, respectively. We also
compare the numerical simulation results of the lattice model with
that of the continuum model. The width of quasi-bound state is
explained by the Fano's formulism.

\textit{Noted added.} While preparing this manuscript, we were aware
of preprint\cite{b6} that has some overlapping with this paper.

\textit{Acknowledgement} This work is partially supported by the
National Natural Science Foundation of China (Grant nos.
10574119). The research is also supported by National Key Basic
Research Program under Grant No. 2006CB922000. Jie Chen would like
to acknowledge the funding support from the Discovery program of
Natural Sciences and Engineering Research Council of Canada under
Grant No. 245680.

\end{document}